\documentclass[traditabstract]{aa} 

\usepackage{graphicx, subfigure}
\usepackage[breaklinks=true]{hyperref} 
\hypersetup{colorlinks=true,citecolor=blue}

\begin{document} 
\title{Testing particle trapping in transition disks with ALMA}
\author{P.~Pinilla\inst{1}, N.~van~der~Marel\inst{1}, L.~M.~P\'erez\inst{2, 3}, E.~F.~van Dishoeck\inst{1,4}, S.~Andrews\inst{5}, T.~Birnstiel\inst{5}, G.~Herczeg\inst{6}, K.~M.~Pontoppidan\inst{7}, T.~van Kempen\inst{1}}
\institute{Leiden Observatory, Leiden University, P.O. Box 9513, 2300RA Leiden, The Netherlands\\
              \email{pinilla@strw.leidenuniv.nl}
              \and
              National Radio Astronomy Observatory, P.O. Box O, Socorro NM 87801, USA
              \and
              Jansky Fellow
              \and
              Max-Planck-Institut f\"{u}r Extraterrestrische Physik, Giessenbachstrasse 1, 85748, Garching, Germany
              \and
               Harvard-Smithsonian Center for Astrophysics, 60 Garden Street, Cambridge, MA 02138, USA
               \and
               Kavli Institute for Astronomy and Astrophysics, Peking University, Yi He Yuan Lu 5, Haidian District, Beijing 100871, China
               \and
              Space Telescope Science Institute, 3700 San Martin Drive, Baltimore, MD 21218, USA 
}  
\date{Received  03 June 2015 / Accepted 08 September 2015}
\authorrunning{P.~Pinilla}


\abstract{Some protoplanetary disks show evidence of inner dust cavities. Recent observations of gas and dust of these so-called transition disks 
support the hypothesis that these cavities originate from particle trapping in pressure bumps. We present new Atacama Large Millimeter/submillimeter Array (ALMA) continuum observations at 336~GHz of two transition disks, SR21 and HD~135344B. In combination with previous ALMA observations from Cycle~0 at 689~GHz, we compare the visibility profiles at the two frequencies and calculate the spectral index ($\alpha_{\rm{mm}}$). The observations of SR~21 show a clear shift in the visibility nulls, indicating radial variations of the inner edge of the cavity at the two wavelengths. Notable radial variations of the spectral index are also detected for SR~21 with values of $\alpha_{\rm{mm}}{\sim}3.8-4.2$ in the inner region ($r\lesssim35$~AU) and $\alpha_{\rm{mm}}{\sim}2.6-3.0$ outside. An axisymmetric ring (which we call the ring model) or a ring with the addition of an azimuthal Gaussian profile, for mimicking a vortex structure (which we call the vortex model), is assumed  for fitting the disk morphology. For SR~21, the ring model better fits  the emission at 336~GHz, conversely the vortex model better fits the 689~GHz emission. For HD~135344B, neither a significant shift in the null of the visibilities nor radial variations of $\alpha_{\rm{mm}}$ are detected.   Furthermore, for HD~135344B, the vortex model fits both frequencies better than the ring model. However, the azimuthal extent of the vortex increases with wavelength, contrary to model predictions for particle trapping by anticyclonic vortices. For both disks, the azimuthal variations of $\alpha_{\rm{mm}}$ remain uncertain to confirm azimuthal trapping. The comparison of the current data with a generic model of dust evolution that includes planet-disk interaction suggests that particles in the outer disk of SR~21 have grown to millimetre sizes and have accumulated in a radial pressure bump, whereas with the current resolution there is not clear evidence of radial trapping in HD~135344B, although it cannot be excluded either.}


\maketitle

\section{Introduction}     \label{introduction}

Recent observations of transition disks (disks with inner dust cavities)  suggest that their structures may originate from particle trapping in pressure bumps \citep[e.g.][]{marel2013, zhang2014}. Particle accumulation in pressure maxima has been suggested to solve the problem of rapid inwards drift of particles \citep[e.g.][]{Weidenschilling1977, Brauer2008}, implying that planetary embryos may form in localised environments \citep[e.g.][]{Klahr1997, Johansen2007}. Pressure bumps may occur because of the presence of one or multiple planets in the disk  \citep[e.g.][]{dodson2011, pinilla2012, pinilla2015, zhu2012}, but other phenomena, such as dead zones may also create pressure traps and explain their structures \citep[e.g.][]{regaly2011, flock2015}. Observations of transition disks reveal that dust cavities  appear to be smaller at shorter wavelengths  \citep[e.g.][]{dong2012, garufi2013}. This spatial segregation of small and mm-sized grains is a natural consequence of particle trapping in the planet-disk interaction scenario \citep[e.g.][]{rice2006, dejuanovelar2013}.

The spectral index $\alpha_{\rm{mm}}$ ($F_\nu\propto\nu^{\alpha_{\rm{mm}}}$) provides information about the particle size in protoplanetary disks \citep[see][for a review]{testi2014}. For (sub-) micron-sized dust, as found  in the interstellar medium, $\alpha_{\rm{mm}}$ is expected to have values of $\gtrsim 3.5-4.0$ \citep[e.g.][]{Finkbeiner1999}. When dust grows to millimetre sizes, $\alpha_{\rm{mm}}$ is expected to decrease \citep{draine2006, ricci2010}. Radial increases in $\alpha_{\rm{mm}}$ (on 100 AU scales) have been found for individual disks without cavities \citep[e.g.][]{Guilloteau2011, perez2012}, consistent with radial drift. In contrast, the inner region of transition disks is depleted of large grains, so that $\alpha_{\rm{mm}}$  would decrease with radius from the central star \citep{pinilla2014}. However, these radial variations for $\alpha_{\rm{mm}}$  in transition disk have not been spatially resolved to date. 

\begin{figure*}
 \centering
  	\includegraphics[width=16.0cm]{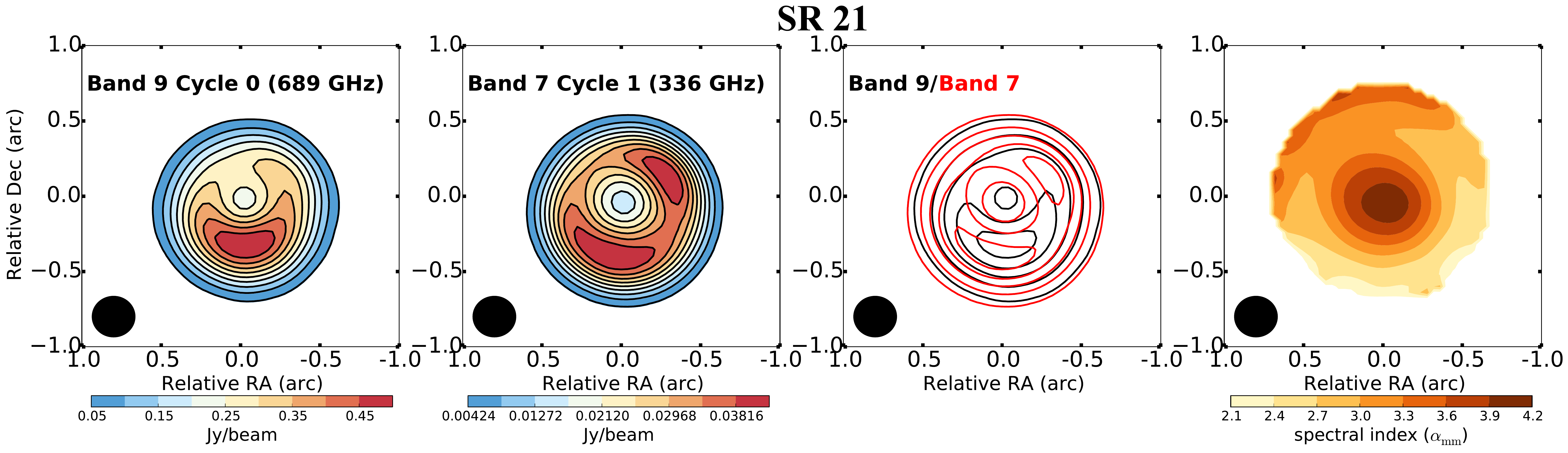}\\ 
	\includegraphics[width=16.0cm]{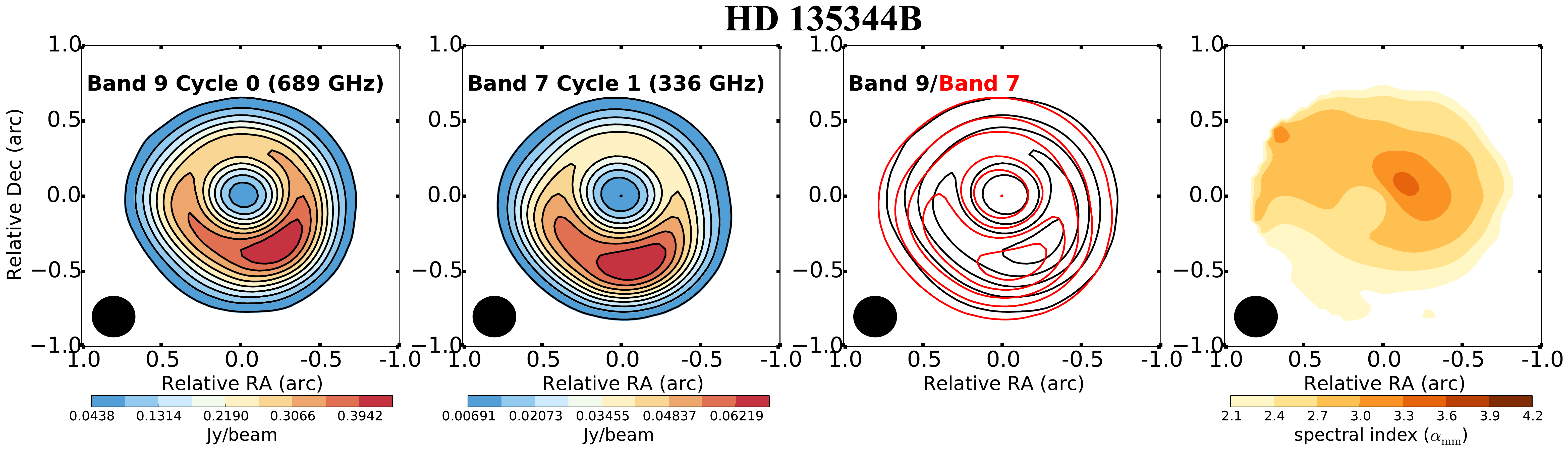} 
\caption{{\small ALMA observations of dust continuum emission for SR~21 (top panels) and HD~135344B (bottom panels). The left columns show the Band 9 (689 GHz) and Band 7 (336 GHz) continuum, with the contour overlaid at 10,20,...,90\% of the peak. The third column shows the overlay of the contours 
with a step of 20\% of the peak (black is Band 9, red is Band 7). The right column shows the resolved $\alpha_{\rm{mm}}$ in colour contours. The beam size is indicated in each plot.}}
   \label{ALMA_obs}
\end{figure*}

In this paper, we combine observations from  ALMA~Cycle~0 at 689~GHz (${\sim}$450~$\mu$m) and Cycle~1 at 336~GHz (${\sim}$~870$\mu$m) of SR~21 and HD~135344B (also known as SAO~206462) to compare the morphological structures at the two frequencies  and calculate $\alpha_{\rm{mm}}$ throughout the disk. In addition, we compare the results with a generic model of particle trapping and dust growth in a pressure bump induced by an embedded planet, and place the results in the context of other transition disks.

\section{Targets and observations}     \label{targets_observations}
SR~21 is a G3 star  located in the Ophiuchus star-forming region at a distance of $d=120$~pc \citep{loinard2008}. The disk was identified as a transition disk by its mid-infrared SED, with a cavity radius of ${{\sim}}18$~AU \citep{brown2007}. Observations with the Submillimeter Array (SMA) at $850~\mu$m confirmed a ${\sim}35$~AU cavity and suggested an azimuthal disk asymmetry \citep{brown2009, andrews2011}, which was much more clearly seen with ALMA observations in Cycle~0 \citep{perez2014}. 
While $450~\mu$m imaging with ALMA shows strong depletion of large mm-sized grains, by a factor of $10^{3}$ or more in the inner region \citep{marel2015}, H-band scattered light imaging indicates that the small grains are much less depleted, perhaps by as little as a factor of 10  \citep{follette2013}.

HD~135344B is an F4 star located in the Sco~OB2-3 star-forming region at a distance of $d=140$~pc \citep{boekel2005}. A cavity radius of ${\sim}45$~AU was also identified by its SED \citep{brown2007} and confirmed by SMA observations  \citep{brown2009}. One of the most intriguing characteristics of this disk is its spiral arms observed in scattered light images  \citep{muto2012, garufi2013}. The observations with VLT/NACO did not show signs of a gap in small dust grains down to 28~AU radius \citep{garufi2013}.

\begin{table}
\caption{Observation properties}
\label{tbl:obsproperties}
\centering   
\tabcolsep=0.055cm          
\begin{tabular}{c|| c | c | c | c | c  | c | c}
\hline
\hline
{\footnotesize Target}&{\footnotesize Band}&{\footnotesize $\nu$}&{\footnotesize $\Delta \nu$}&{\footnotesize $F_{\rm{peak}}$}&{\footnotesize $F_{\rm{total}}$}&{\footnotesize $\sigma$}&{\footnotesize Date}\\
&&{\footnotesize (GHz)}&{\footnotesize (GHz)}&{\footnotesize (mJy)}&{\footnotesize (mJy)}&{\footnotesize (mJy)}&\\
\hline
{\footnotesize SR~21}&$7$&$336$&$3.2$&$42.4$&349&$0.08$&{\footnotesize 26/07/2014}\\ 
&$9$&$689$&$7.5$&$500$&2877&$2.0$&{\footnotesize 18/07/2012}\\ 
\hline
{\footnotesize HD~135344B}&$7$&$336$&$3.2$&$69.1$&636&$0.13$&{\footnotesize 27/07/2014}\\ 
&$9$&$689$&$7.5$&$438$&3360&$2.5$&{\footnotesize 05/07/2012}\\ 
\hline
\end{tabular}
\end{table}

Our observations are from ALMA Cycle~0 program 2011.0.00724.S (P.I. P{\'e}rez) and Cycle~1 program 2012.1.00158.S (P.I. van Dishoeck), taken in Band~9 and Band~7, respectively. The details of the calibration are described in \cite{perez2014} and van der Marel et al. (submitted). For both observation sets, the images were obtained using Briggs weighting with a robust parameter of 0.5 and and the beam is set to be the same at both bands by the post-processing of the images, which is the best possible beam size corresponding to both the Band 7 and Band 9 observations (0.27''). The details of the images are given in Table \ref{tbl:obsproperties}.

As the data are taken almost two years apart, the centre of the 2014 data was shifted before overlaying the images to compensate for the proper motion of the stars. The proper motion is -12,-24~mas~yr$^{-1}$ for SR~21 \citep{Makarov2007} and -20,-24~mas~yr$^{-1}$ for HD~135344B \citep{Hog2000}. Figure~\ref{ALMA_obs} shows the continuum maps and their overlay.  The shape and contrast of the asymmetry look similar between Band 7 and 9 for HD~135344B. In contrast, SR~21 looks more symmetric in Band~7. There is an azimuthal shift in the continuum overlay for both sources. The position of the gain calibrator (the same for both observations) shifts $\lesssim$0.1~mas, meaning that the azimuthal shift is not due to calibration. With respect to the SMA images in 2006/2007 \citep{brown2009}, the asymmetries are also shifted.

\section{Data analysis}     \label{data_obs}

\begin{table*}
\caption{Best-fit parameters the disk morphology models of ring or ring+vortex.}
\centering   
\tabcolsep=0.2cm
\subtable[Ring model]{                   
\begin{tabular}{c||ccccc}       
\hline 
\hline 
{\small Target}& {\small $\nu$}&{\small $\chi^2$}& {\small $F_R$}& {\small $r_R$}&{\small $\sigma_R$}\\
& {\small (GHz)}&&{\small ($\mu$m)}& {\small (AU)}&{\small (AU)}\\
\hline
& 336 & {\tiny 2.07}&{\tiny $0.71$} & {\tiny $41$} & {\tiny $12$} \\
SR~21 & & &  \\ 
&689 & {\tiny 1.39} &{\tiny $5.91$} & {\tiny $36$} & {\tiny $15$} \\
\hline
& 336 & {\tiny 3.75} &{\tiny $0.84$} & {\tiny $61$} & {\tiny $19$} \\
HD~135344B & & &  \\ 
&689 & {\tiny 1.18} &{\tiny $5.72$} & {\tiny $60$} & {\tiny $18$} \\
\hline
\end{tabular} } 
\subtable[Ring+Vortex model] {
\begin{tabular}{c||cccccccccc}
\hline
\hline
{\small Target}& {\small $\nu$}&{\small $\chi^2$}&{\small $F_R$}& {\small $r_R$}&{\small $\sigma_R$}&{\small $F_V$}&{\small $r_V$}&{\small $\theta_V$}&{\small $\sigma_{R, V}$}&{\small $\sigma_{\theta, V}$}\\
& {\small (GHz)}&&{\small ($\mu$m)}& {\small (AU)}&{\small (AU)}&{\small ($\mu$m)}& {\small (AU)}&{\small ($^{\circ}$)} & {\small (AU)} & {\small (AU)}\\
\hline
& & & & & & & & & &  \\
SR~21&689 & {\tiny 1.06} &{\tiny $4.72$}&{\tiny $35$}&{\tiny $14$}&{\tiny $4.0$}&{\tiny $46$}&{\tiny $178$}&{\tiny $14$}&{\tiny $40$}\\
& & & & & & & & & &  \\
\hline
& 336 & {\tiny 1.52} &{\tiny $0.70$}&{\tiny $70$}&{\tiny $14$}&{\tiny $0.88$}&{\tiny $43$}&{\tiny $172$}&{\tiny $16$}&{\tiny $53$} \\
HD~135344B & & & & & & & & & &  \\
&689 & {\tiny 1.05} & {\tiny $5.24$} & {\tiny $65$} & {\tiny $16$} & {\tiny $7.0$} & {\tiny  $42$} & {\tiny $194$} & {\tiny $7.0$} & {\tiny $47$}  \\
\hline
\end{tabular} 
}
\label{best_parameters}
\tablefoot{{\small All data of Band~9 are identical than \cite{perez2014}. The parameters for SR~21 in Band~7 of the ring+vortex model are omitted because of the unphysical results. The errors from the MCMC calculations are much smaller than the spatial uncertainty from the observations (and therefore omitted), which is ${\sim}10\%$ of the beam size (i.e. ${\sim}$3~AU for SR~21 and ${\sim}$4~AU for HD~135344B).}}  
\end{table*}

\subsection{Visibilities and disk morphology} 

\begin{figure}
   \centering
   \includegraphics[width=8.0cm]{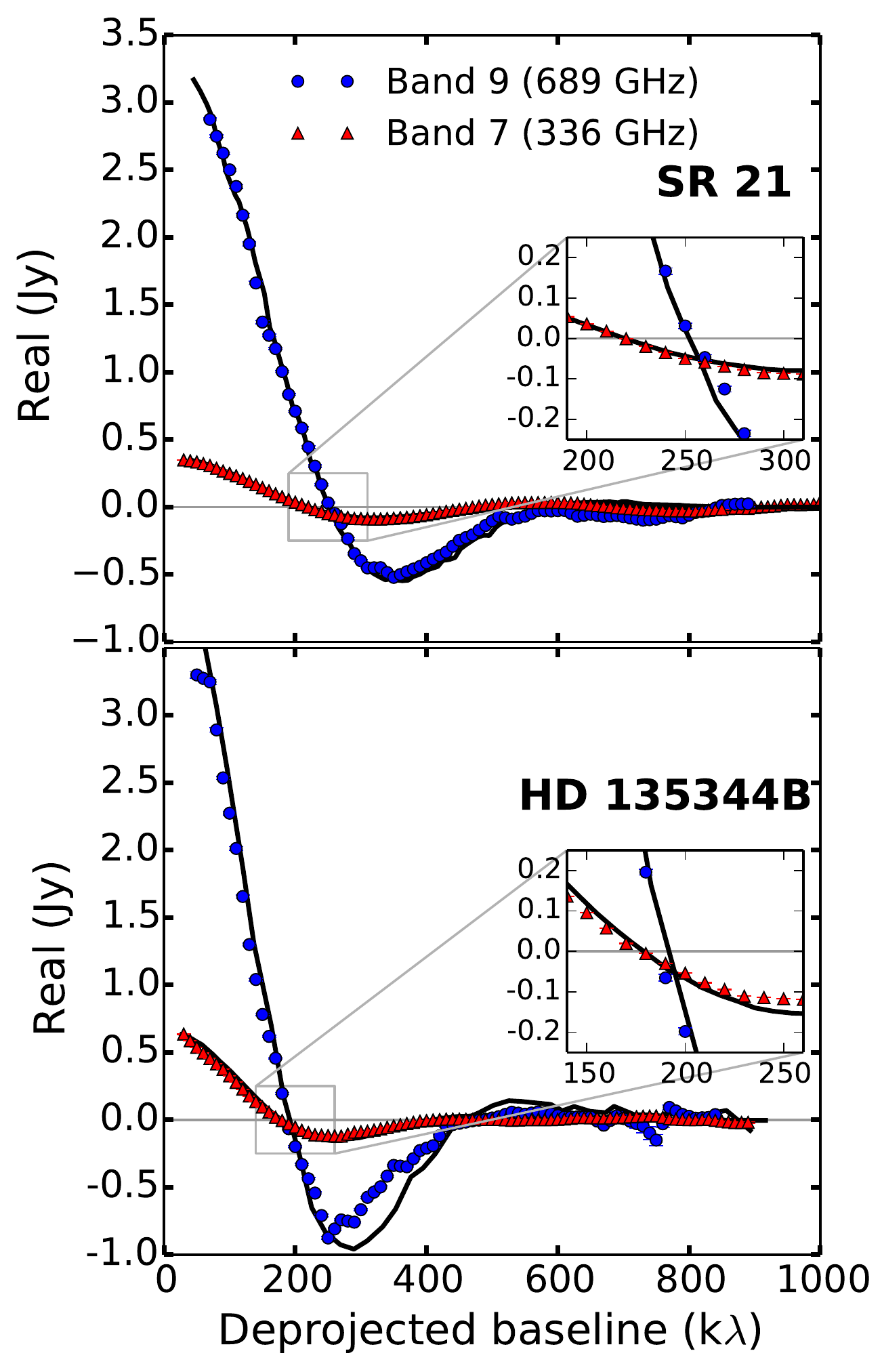}
   \caption{{\small Real part of the visibilities at 689~GHz (${\sim}$450~$\mu$m, Band~9) and at 336~Gz (${\sim}$870~$\mu$m, Band 7) for SR~21 (top panel) and HD~135344B (bottom panel). The plot includes error bars that are of the size of the plotting points. At both wavelengths, the models with the best-fit parameters are over-plotted (solid line), which is the ring model  in the case of SR~21, and the vortex model for HD~135344B.} }
   \label{visi_models_obs}
\end{figure}

The real part of the visibilities at both frequencies is shown in Fig. \ref{visi_models_obs}. These are extracted from the continuum data and de-projected using $i=15^{\circ}$ and position angle P.A.=14$^{\circ}$ for SR~21, and $i=20^{\circ}$ and P.A.=63$^{\circ}$, as derived in \cite{pontoppidan2008} and \cite{marel2015}. The data are binned by taking the mean of the available data points in bins of 20 k$\lambda$, with a minimum of five data points per bin. In Fig.~\ref{visi_models_obs}, we show the real part of the visibilities for the two bands. The length of the projected baseline where the visibilities cross zero (the null) indicates the location inner edge of the cavity in a given bandpass (smaller values for the null mean that the cavity inner edge is further out for the same disk properties). The nulls for SR~21 are ${\sim}$220~k$\lambda$ at Band 9 and ${\sim}$250~k$\lambda$ at Band~7. For HD~135344B the null is almost at the same position at both wavelengths (${\sim}$190~k$\lambda$). The imaginary part of the visibilities are presented in Appendix~\ref{appendix_a}.

To fit the visibility profiles and constrain the structure of both disks, we used the same morphological models described in \cite{perez2014}. One model is a ring-like emission described by $F(r,\theta)=F_R~e^{-(r-r_R)^2/2\sigma_R^2}$, where $r_R$ is the location of the peak of the ring emission, $F_R$ the flux density at $r_R$, and $\sigma_R$ the ring width (ring model). The other model assumes a ring together with an azimuthal Gaussian profile to mimic a vortex structure \citep{lyra2013}. The vortex is described by $F(r,\theta)=F_V~e^{-(r-r_V)^2/2\sigma_{r,V}^2}e^{-(\theta-\theta_V)^2/2\sigma_{\theta,V}^2}$, with $r_V$ and $\theta_V$ being the radius and P.A. at the peak of the vortex, $F_V$ the flux density at $(r_V,\theta_V)$, and $\sigma_{\theta,V},\sigma_{r,V}$ is the width of the vortex in the radial and azimuthal direction respectively (vortex model). The parameters of the best-fit model are found by minimising $\chi^2$ using the same Markov Chain Monte Carlo (MCMC) approach implemented by \cite{perez2014}. The results are summarised in Table~\ref{best_parameters}. The errors from the MCMC calculations are much smaller than the spatial uncertainty from the observations, which is typically ${\sim}\,10\%$ of the beam size (i.e. ${\sim}\,$3~AU for SR~21 and ${\sim}\,$4 for HD~135344B) for the high signal to noise of the data (Table~\ref{tbl:obsproperties}). For the MCMC calculations, the inclination and P.A. are fixed to the values found by \cite{perez2014} for both disks. The model including a vortex did not converge to a physical solution for SR~21 in Band~7, because the azimuthal extension of the vortex covers more than $2\pi$. In summary, the SR 21 data are best fit with a ring model, while the HD 135344B data are best fit with a vortex model (see Fig.~\ref{visi_models_obs}). The residuals obtained by subtracting the best-fit model from the data in Band~7 are shown for both targets in Appendix~\ref{appendix_c}.

In the case of SR 21, the best-fit model indicates that larger grains, as traced at longer wavelengths by Band 7, are more spatially confined than the smaller grains traced at shorter wavelengths by Band 9 ($\sigma_R{\sim}12$~AU at 870~$\mu$m vs. $\sigma_R{\sim}15$~AU at 450~$\mu$m). Furthermore, the peak of emission in the ring is at larger radii in Band 7 than in Band 9 ($r_R{\sim}41$~AU at  870~$\mu$m vs. $r_R{\sim}36$~AU at 450~$\mu$m). The location of the inner edge of the cavity at the two frequencies varies from  ${\sim}29$~AU in Band~9 to ${\sim}35$~AU in Band~7, which together with the wider ring,  is consistent with the shift of the null of the visibilities (Fig.~\ref{visi_models_obs}).  For HD~135344B, the vortex model is consistently the best-fit model for both wavelengths. The inner edge of the cavity  is almost constant in the two bands, as shown in the visibility nulls  (Fig.~\ref{visi_models_obs}). However, the width of the vortex in both the radial and azimuthal directions is significantly larger in Band 7, and its aspect ratio ($\sigma_{\theta,V}/\sigma_{r,V}$) decreases from 7.1 in Band 9 to 3.3 in Band 7.

\subsection{Spectral index} 

The position-dependent spectral index (right panels of Fig.~\ref{ALMA_obs}) is calculated using both bands as $\alpha_{\rm{mm}} = \ln\frac{F_{\nu_{B7}}}{F_{\nu_{B9}}} /\ln\frac{\nu_{B7}}{\nu_{B9}}$. All data with $F<5\sigma$ are excluded. The dust emission within the cavity is detected at a high level of significance (S/N${\sim}320\sigma$ and ${\sim}250\sigma$ in Band~7, and ${\sim}120\sigma$ and ${\sim}80\sigma$ in Band~9, for SR~21 and HD~135344B, respectively), so $\alpha_{\rm{mm}}$ can be accurately computed in the cavity separately from the dust ring. The uncertainty of $\alpha_{\rm{mm}}$ is calculated with error propagation from the observational uncertainty, which includes the calibration uncertainties (${\sim}$20\% and ${\sim}$10\% in Band 9 and 7, respectively) and the noise level or rms  ($\sigma$) of the observations. With the high signal-to-noise of the data, the uncertainty in the spectral index $\sigma_{\alpha_{\rm{mm}}}$ at a given position is dominated by the calibration uncertainties and approximates to $\approx(\ln \nu_1/\nu_2)^{-1}\sqrt{(0.2^2+0.1^2)}\approx0.3$, in both cases. However, for the relative spatial changes in  $\alpha_{\rm{mm}}$,  the systematic calibration uncertainty does not contribute since it is constant across the field, and changes in $\alpha_{\rm{mm}}$ inside and outside the dust cavity have high significance with maximum values of $\sigma_{\alpha_{\rm{mm}}}$ of $\approx 0.01$. SR~21 shows significant radial variations of $\alpha_{\rm{mm}}$, with values of $\alpha_{\rm{mm}}{\sim}3.8-4.2$ within ${\sim}$0.3'' radius (${\sim}$35~AU) and $\alpha_{\rm{mm}}{\sim}2.6-3.0$ outside ($\Delta \alpha_{\rm{mm}}\gtrsim 120\sigma_{\alpha_{\rm{mm}}}$). This implies a lack of mm-grains in the inner region ($r\lesssim35~$AU), and is consistent with accumulation of mm-grains in a localised region in the outer disk.  For HD~135344B, no significant radial variations for $\alpha_{\rm{mm}}$ are found, and the values remain within a range of $2.6-3.2$. Although Fig.~\ref{ALMA_obs} shows azimuthal variations of $\alpha_{\rm{mm}}$ for both sources, these depend considerably on how the two images are overlaid in the calculation of  $\alpha_{\rm{mm}}$ (see Appendix~\ref{appendix_b}), and therefore any azimuthal variation within a range of $\alpha_{\rm{mm}}{\sim}eq 2.6-3.2$ remains uncertain.  

\section{Theoretical predictions of particle trapping} \label{theory}

 Dust trapping, in radial and azimuthal pressure bumps, depends on the coupling of the particles to the gas, pressure gradient, and  disk turbulence. The dimensionless stopping time, Stokes number, quantifies the coupling of the particles and it is defined in the mid-plane (and assuming a Gaussian vertical profile for the gas density) as St$=a\rho_s\pi/2\Sigma_g$ \citep[with $\rho_s$ being the volume density of a grain, with typical values of ${\sim}1\rm{g~cm}^{-3}$,][and $\Sigma_g$ the local gas surface density]{blum2008}. Particles with $\rm{St}{\sim}1$ feel the strongest gas drag and  therefore they move much faster to the regions of pressure maxima \citep[e.g.][]{birnstiel2010, pinilla2012}. As the gas is turbulent, however,  it is expected that the dust is turbulently mixed by the gas. The dust diffusion  prevents the concentration of all the particles with $\rm{St}{\sim}1$ inside pressure traps. The drift of particles (and hence the trapping) is efficient for particles with $\rm{St}\gtrsim\alpha_{\rm{visc}}$, where $\alpha_{\rm{visc}}$ is a dimensionless number that quantifies the disk viscosity \citep{shakura1973}. From the combination of radial drift and dust diffusion,  it is expected that particles with $\rm{St}{\sim}1$ are more concentrated at the pressure maximum than $\rm{St}<1$. For example, the models of dust trapping by a vortex predict that larger grains would be more azimuthally concentrated in the centre of the vortex \citep{birnstiel2013, lyra2013}, as observed in IRS~48 \citep{marel2015b} and HD~142527 \citep{casassus2015}. Although turbulence can change the  broadness of the dust concentration inside pressure bumps, it does not change the location of  pressure maxima, i.e. the mean radial/azimuthal location of the dust concentration.
 
\begin{figure}
    \centering
  	\includegraphics[width=8.0cm]{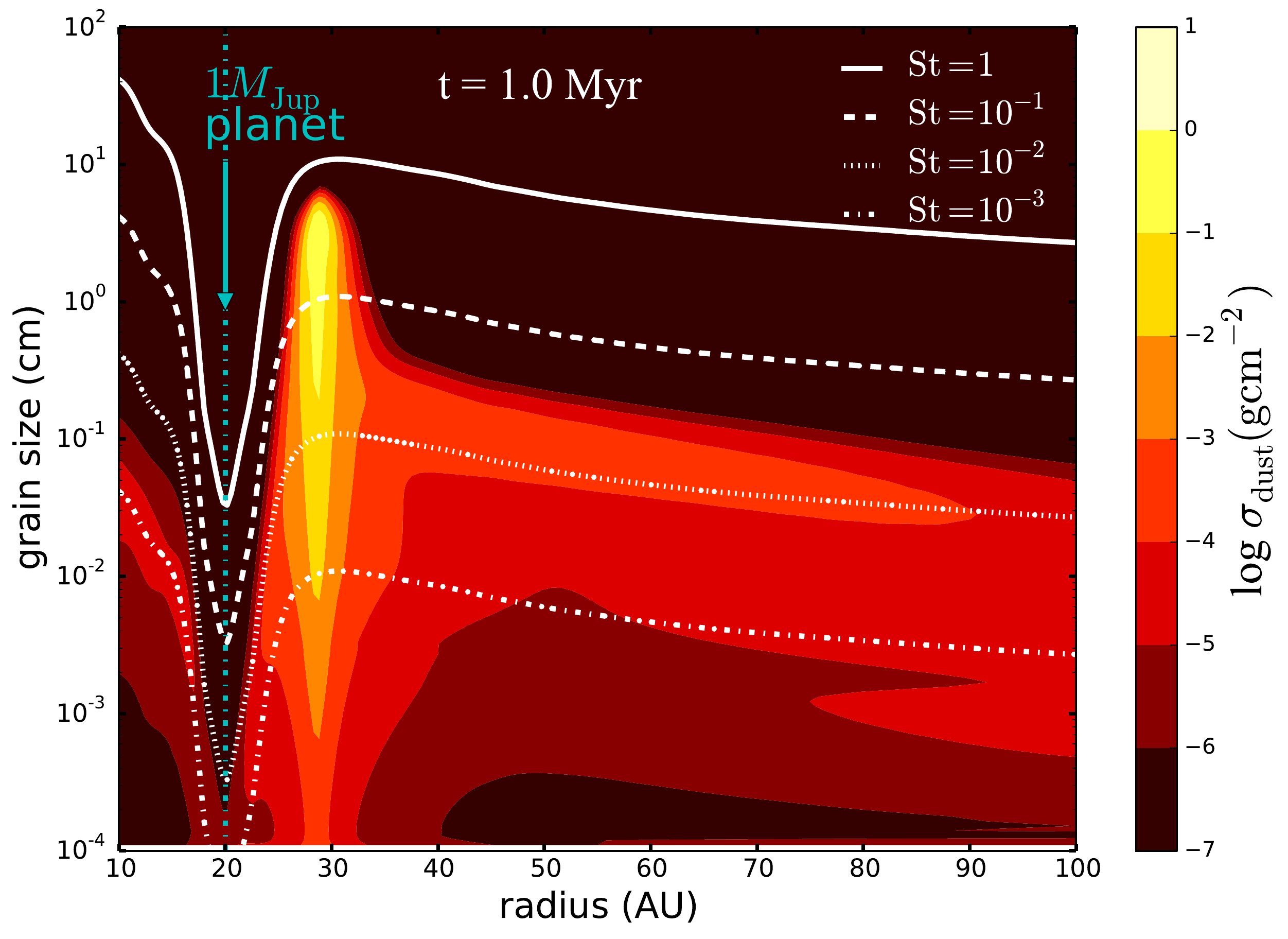}\\
	 \includegraphics[width=8.0cm]{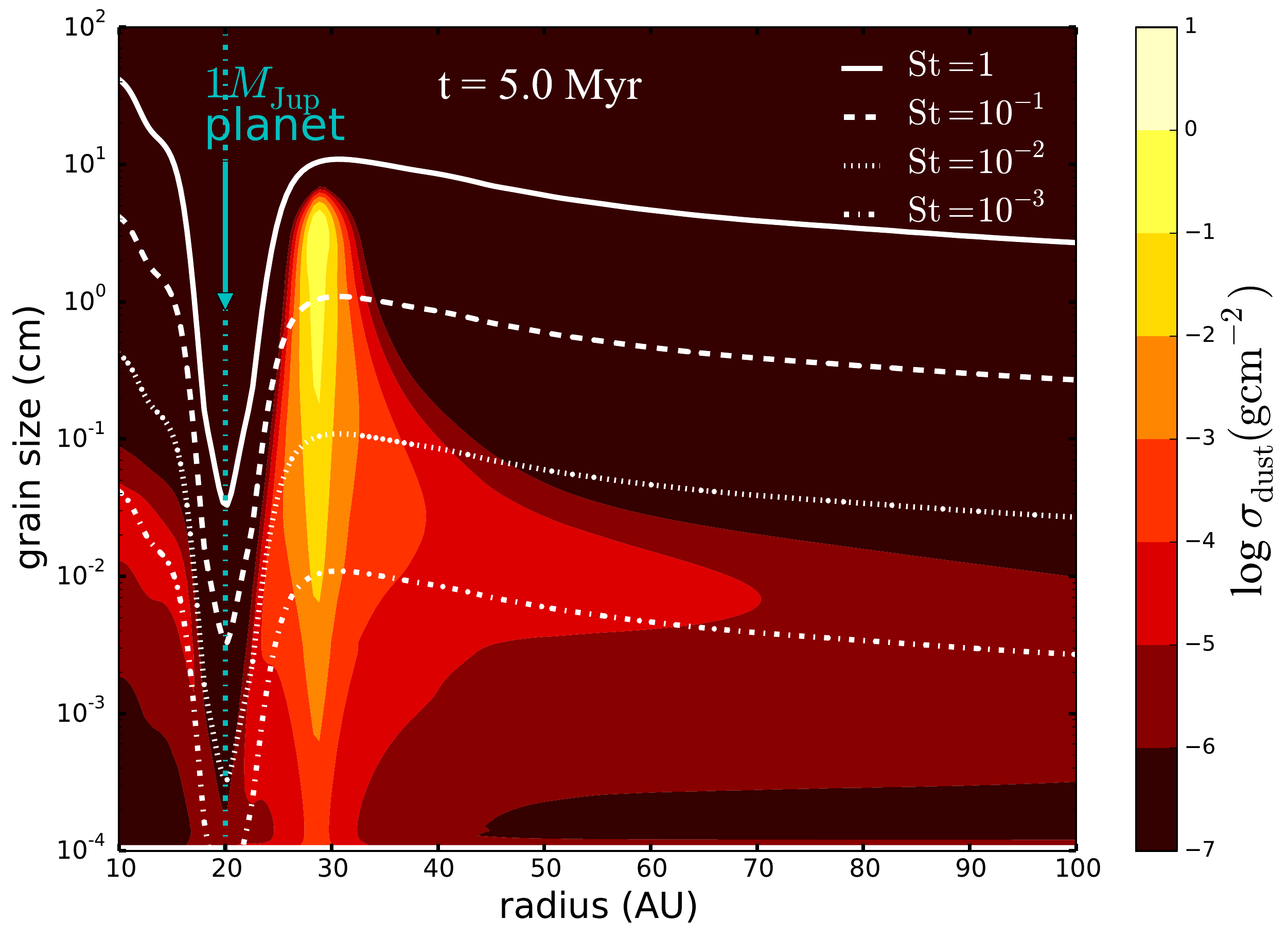}
   \caption{{\small Dust density distribution when $1~M_{\rm{Jup}}$ is embedded at 20~AU after 1 and 5~Myr of evolution (top and bottom panel respectively). The lines represent particle sizes with different Stokes numbers (St=$[10^{-3}, 10^{-2}, 10^{-1}, 1]$), which are proportional to the gas surface density profile.}} 
   \label{model_comparison1}
\end{figure}
 
\begin{figure}
 \centering
  	\includegraphics[width=8.0cm]{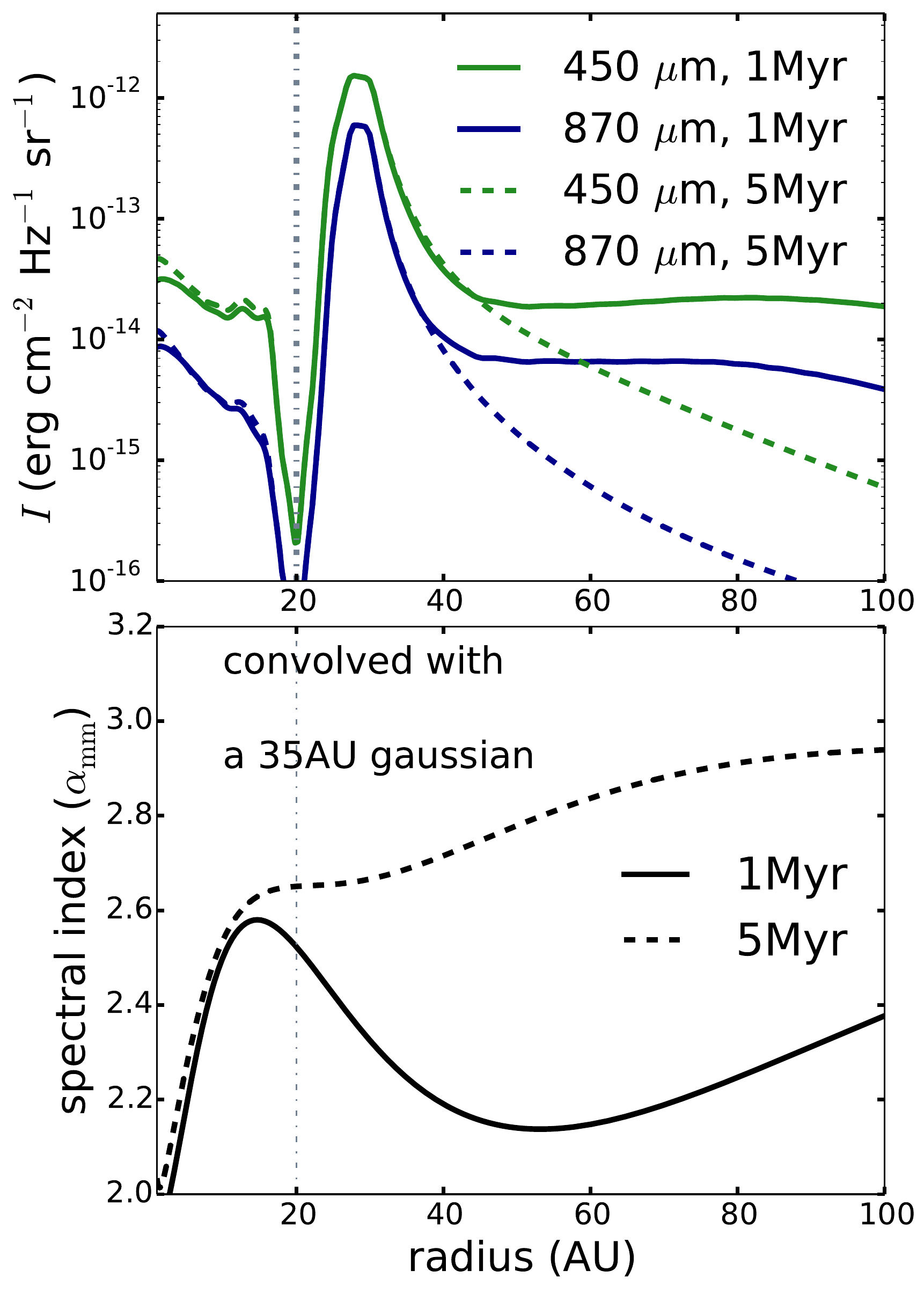}
\caption{{\small Top panel: intensity radial profiles at 450 and  870~$\mu$m obtained from the dust density distributions after 1 and 5 Myr of evolution. Bottom panel: radial profile of $\alpha_{\rm{mm}}$ calculated from the intensity profiles at 450 and 870~$\mu$m,  which is convolved  with 35 AU Gaussian beam.}} 
   \label{model_comparison2}
\end{figure}

 Figure~\ref{model_comparison1} shows the model predictions for  particle trapping triggered by planet-disk interaction. The gas surface density is obtained from hydrodynamical simulations of a $1~M_{\rm{Jup}}$ planet embedded in the disk at 20~AU distance from the star as in \cite{pinilla2012}. The initial gas surface density and temperature are assumed to be a power law, and the disk viscosity is taken to be $\alpha_{\rm{visc}}=10^{-3}$, as found for typical values of disk turbulence in simulations of MRI-active disks \citep[e.g.][]{dzyurkevich2010}. The dust density distributions are obtained from dust evolution models that include the dynamics and coagulation of dust particles \cite{birnstiel2010}. When a planet opens a gap in the disk, a pressure bump is formed at the outer edge of the gap, and large particles drift to the pressure maximum located in this case at ${\sim}~30$~AU.  The particle sizes with different Stokes number (St=$[10^{-3}, 10^{-2}, 10^{-1}, 1]$) are shown to illustrate how the radial concentration becomes narrower for particles with higher St. The maximum grain size inside the trap is limited by dust fragmentation, which depends on  $\alpha_{\rm{visc}}$ \citep[e.g.][]{birnstiel2010}. For higher turbulence, the maximum grain size that particles can reach before fragmentation would be lower, causing the particles to easily escape from the bumps and then drift radially inwards. For more turbulent disks, a stronger pressure bump (with e.g. a more massive planet) would be required for trapping to occur \citep[e.g.][]{pinilla2015}. Figure~\ref{model_comparison2} shows the intensity profile at  at 450 and 870~$\mu$m. The emission profile is only slightly narrower at 870~$\mu$m since the range of particles sizes that are traced is comparable. Both the dust density distributions and the intensity profiles become narrower at longer times of evolution.
 
Because of particle trapping, the  region close to the planet is empty of mm/cm particles, and, therefore, $\alpha_{\rm{mm}}$ is expected to increase close to the location of the gap carved by the planet.  Figure.~\ref{model_comparison2} also shows the expected radial profiles of $\alpha_{\rm{mm}}$ calculated from the models at 450 and 870~$\mu$m and convolved with a 2D Gaussian profile of 35 AU diameter. In the inner part ($r\lesssim10$~AU),  $\alpha_{\rm{mm}}$ drops because of the presence of the mm-particles. Because there is no total filtration of particles at the outer edge of the gap, small grains still flow through the gap and grow again to mm-sizes in the inner region ($r\lesssim10~$AU). One way to prevent the presence of mm-grains is to increase the mass of the planet, and completely filter all particle sizes \citep{zhu2012, pinilla2012, pinilla2015}.
 
At early times of dust evolution (${\sim}1$~Myr), mm/cm grains are still distributed in the entire outer disk ($r\gtrsim30$~AU) and radial variations of $\alpha_{\rm{mm}}$ are detected after convolution. This is not the case at 5~Myr, since the concentration of mm/cm particles becomes much narrower compared to the spatial resolution and hence the potential radial variation of $\alpha_{\rm{mm}}$ is smeared out and is not detected.

\section{Discussion}
With the new ALMA observations of SR~21 and HD~135344B, it is possible to test whether radial/azimuthal trapping is the cause of disk structures (cavities and asymmetries). 
Observationally,  radial trapping can be tested by analysing the location of the null in the real part of the visibilities at different wavelengths because, in the case of trapping, the mm-emission is expected to show a wider ring at shorter wavelengths. Thus,  the inner edge of the ring (or dust cavity) should be located closer to the star at shorter wavelengths (Sect.~\ref{theory}). Additional insights can be gained by analysing the wavelength-dependent morphology and the position-dependent spectral index.

\paragraph{SR~21:} The current observations of SR~21 at 336 and 689~GHz suggest that the disk morphology at longer wavelengths is better described by a ring than by a vortex. 
From observations of $^{12}$CO, the disk mass of SR~21 has been inferred to be ${\sim}12~M_{\rm{Jup}}$, and the average value of the gas surface density close to the cavity is ${\sim}80~\rm{g~cm}^{-2}$ \citep{marel2015}.  Analysis of recent observations of $^{13}$CO  and C$^{18}$O of this disk confirmed the disk mass (van der Marel, submitted). Assuming that at each wavelength, particles with a maximum size of $\lesssim3\lambda$ are traced \citep[e.g.][]{draine2006}, observations at Band 7 and 9 traced particles with $\rm{St}{\sim}2.5\times10^{-3}-5.0\times10^{-3}$ close to the location of the cavity. These particles are still expected to be affected by radial drift and to move towards pressure maxima.

The fact that at a longer wavelength the morphology is  better described by a ring than a vortex is in contradiction with model predictions of trapping by a vortex. Thus, the observed azimuthal structure is likely not caused by a vortex. Moreover, to sustain a long-lived vortex, the radial width of the vortex cannot be much higher than the disk scale height \citep[$\lesssim 2h$,][]{barranco2005} and the fact that the radial width of the vortex in SR~21 is much larger that the disk scale height \citep[${\sim}\times5~h$][]{marel2015}, as found from the morphology fitting of the Band~9 data (Table~\ref{best_parameters}), also disfavours the vortex scenario.

In the radial direction, larger grains traced at 870~$\mu$m are more narrowly concentrated than smaller grains traced at 450~$\mu$m in SR~21, as demonstrated  in the shift of the nulls of the visibilities (see Fig.~\ref{visi_models_obs} and Table~\ref{best_parameters}). The narrower concentration of larger grains is in agreement with particle trapping by a radial pressure bump. From measuring the broadness of the dust concentration inside the trap at different wavelengths, the disk turbulence could be estimated (broader implying higher turbulence)  if the disk temperature and gas surface density profile are well constraint from observations. Further evidence that radial trapping occurs in SR~21 is given by the radial changes of the spectral index $\alpha_{\rm{mm}}$, which increase outwards, as expected from radial trapping models (Fig.~\ref{model_comparison2}). The detection of radial variation of  $\alpha_{\rm{mm}}$ with the current resolution suggests that particle trapping in SR~21 has occurred in $\lesssim1$~Myr because at longer times variations of $\alpha_{\rm{mm}}$ would not be detected (Fig.~\ref{model_comparison2}). The discrepancy between model predictions and observations for the values of $\alpha_{\rm{mm}}$ in the inner part of the disk come from the fact that we do not have total filtration of particles in the outer edge of the gap in our model.
 
Another potential explanation for variations of $\alpha_{\rm{mm}}$ is high optical depth, in which case the physical temperature is expected to be close to the brightness temperature. To reproduce the observed differences in $\alpha_{\rm{mm}}$ with optical depth effects alone, the emission must originate from a small surface area.  Assuming that the emission is distributed in a homogeneous ring whose temperature is equal to the highest  brightness temperature at the peak of emission obtained at the two frequencies ($T_B\approx$32~K in Band~9 and $\approx$13~K in Band~7), the ring must be very narrow ($\lesssim$3 AU in width). This contradicts the observed spatial extent of the emission in both bands (30-40AU for SR~21 and 60-70AU for HD~135344B, see Table~\ref{best_parameters}). Thus, the radial variations of  $\alpha_{\rm{mm}}$ cannot be due to optically thick emission alone. \cite{perez2014} demonstrated that the emission at Band 9 is marginally optically thick at the peak of emission, so it can only trace some of the mass surface density, and hence high angular resolution observations at longer wavelength are necessary to confirm our predictions at the location of the peak.

\paragraph{HD~135344B:} In this case, the model with the best-fit parameters for the disk morphology at the two wavelength is the vortex model  (Table~\ref{best_parameters}). The disk mass inferred from $^{12}$CO, $^{13}$CO, and C$^{18}$O is ${\sim}20~M_{\rm{Jup}}$ \citep[][and submitted]{marel2015}, and the average value of the gas surface density at the location of the dust cavity is similar to that in SR~21 (${\sim}80~\rm{g~cm}^{-2}$). Therefore, the current observations trace particles with similar  Stokes numbers as SR~21. 

From the morphological models, the azimuthal width of the vortex  increases at longer wavelengths in contradiction with predictions of particle trapping by vortices. Similar to SR~21, the radial width of the vortex is too large compared to the scale height of the disk. The origin of the azimuthal asymmetry is inconsistent with a vortex and may be related to the spiral arms observed at scattered light as also suggested by \cite{perez2014}. In particular the bright spiral in the south coincides with the location of the asymmetry in the millimetre \citep[][see also Appendix~\ref{appendix_c}]{garufi2013, quanz2015}.  Indeed, the azimuthal shift of the peak (Fig.~\ref{ALMA_obs}, also observed for SR~21) may be related to the physical rotation of spiral arms, but higher angular resolution observations are needed to confirm this prediction.

In contrast with SR~21, neither a shift of the null of the visibilities nor radial variations of $\alpha_{\rm{mm}}$ are detected for HD~135344B, but this does not exclude radial particle trapping. From the model predictions (Sect.~\ref{theory}), the dust density distribution of mm-grains is more narrowly concentrated at the pressure maximum after 5~Myr than at 1~Myr, and radial variations in $\alpha_{\rm{mm}}$ would not be detected with the current resolution (Fig.~\ref{model_comparison2}). Hence, the fact that radial changes in $\alpha_{\rm{mm}}$ are not detected for HD~135344B can be related to the fact that any instability (e.g. planet) responsible for the trapping  formed earlier (${\sim}$5~Myr ago) in HD~135344B than in SR~21 (${\sim}$1~Myr ago). Another possibility is that trapping happens in more refined structures, such as spiral arms in gravitationally unstable disks, in which case any variations of $\alpha_{\rm{mm}}$ remains unresolved \citep{dipierro2015}.

Besides longer evolution times, other disk and planet parameters, such as planet mass, turbulence, or disk temperature, can also affect the gap shape and thus the radial concentration of mm/cm-sized particles in pressure bumps, leading to a different spatial distribution of small and large grains. 
High contrast asymmetries have been observed in other transition disks and interpreted as vortices \citep[e.g. Oph~IRS~48,][]{marel2013}. Detection of vortices in transition disks may be atypical because several parameters, such as strong turbulence or feedback from dust to the gas, can prevent a vortex from being long-lived \citep[e.g.][]{ataiee2013, zhu2014, raetting2015}.  Even in the case where a planet originally triggered the formation of the vortex, an eccentric orbit or the presence of addition planets can also lead to its rapid destruction \citep{ataiee2015}. 

The current ALMA observations of SR~21 and HD~135344B suggest that anti-cyclonic vortices are not the origin of their low contrast azimuthal asymmetries. Observations at high angular resolution at longer, optically thin wavelengths, which provide information about the distribution of larger grains, will further constrain the origin of the observed dust structures in transition disks.

\begin{acknowledgements}
We are thankful to the anonymous referee, who helped to improve the quality of this manuscript. We are grateful with M.~Benisty and C.~P.~Dullemond for fruitful discussions. Astrochemistry in Leiden is supported by the Netherlands Research School for Astronomy (NOVA), 
by a Royal Netherlands Academy of Arts and Sciences (KNAW) professor prize, 
and by the European Union A-ERC grant 291141 CHEMPLAN.  
T.~B. acknowledges support from NASA Origins of Solar Systems grant NNX12AJ04G. 
This paper makes use of the following ALMA data: ADS/JAO.ALMA\#2012.1.00158.S and \#2011.0.00724.S. 
ALMA is a partnership of ESO (representing its member states), NSF (USA), and NINS (Japan), together with NRC (Canada) and NSC and ASIAA (Taiwan), in cooperation with the Republic of Chile.  The Joint ALMA Observatory is operated by ESO, AUI/NRAO and NAOJ.  

\end{acknowledgements}

\bibliographystyle{aa}
\bibliography{ALMA_paper.bbl}

\appendix

\section{Imaginary part of the visibilities} \label{appendix_a}

\begin{figure}[h!]
 \centering 
   \includegraphics[width=9.0cm]{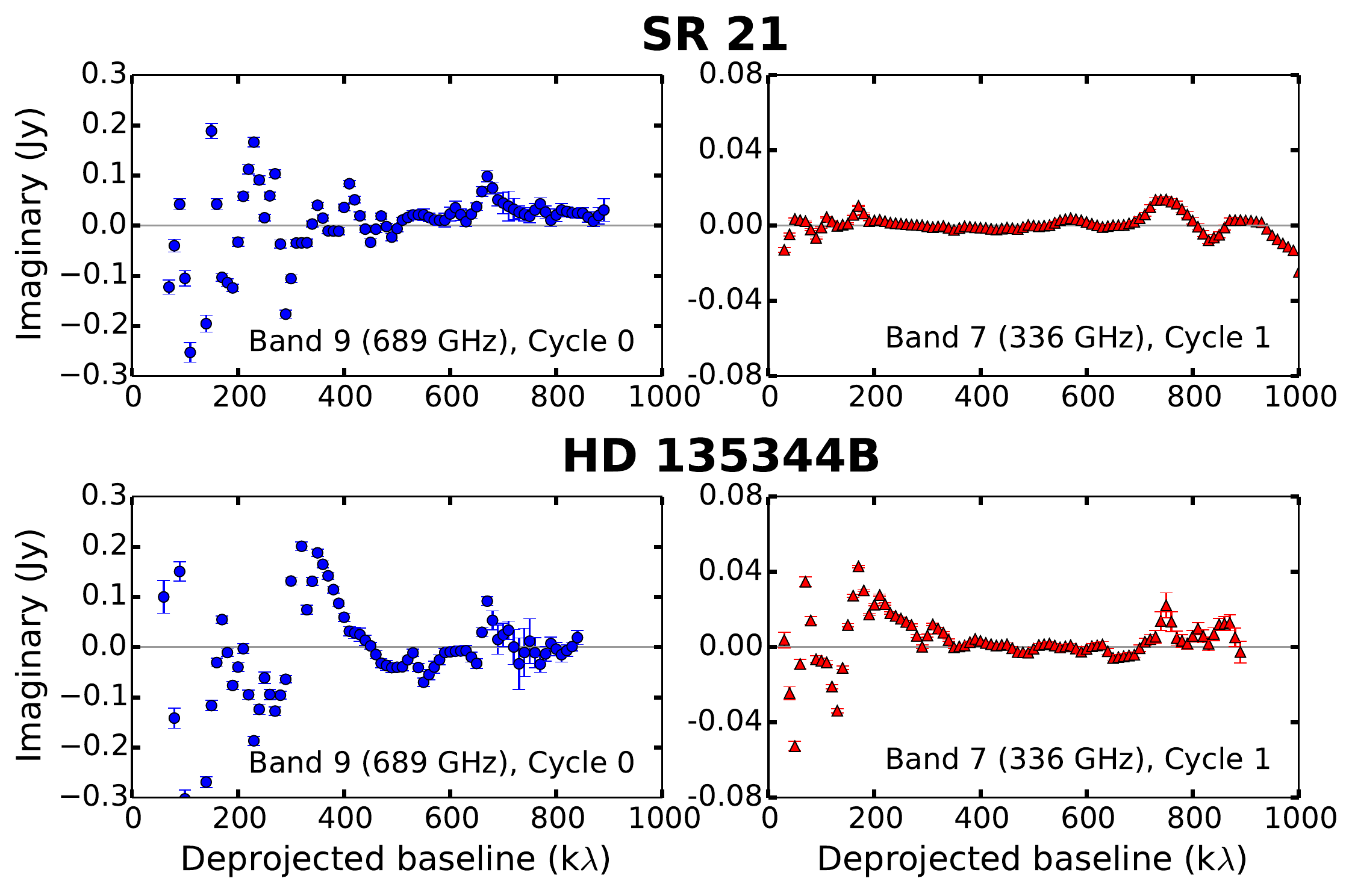}
   \caption{{\small Imaginary part of the visibilities at 689~GHz (${\sim}$450~$\mu$m, Band~9) and at 336~GHz (${\sim}$870~$\mu$m, Band7) for SR~21 (top panel) and HD~135344B (bottom panel).}}
   \label{imaginary_part}
\end{figure}

Figure~\ref{imaginary_part} shows the imaginary part of the visibilities at 689~GHz (${\sim}$450~$\mu$m, Band~9) and at 336~Gz (${\sim}$870~$\mu$m, Band7) for SR~21 and HD~135344B.  Non-zero imaginary visibilities indicate the presence of an azimuthal asymmetry \citep[e.g.][]{isella2013}. For HD~135344B, there are significant variations from zero at both frequencies. For SR~21, the non-zero values are marginal in Band 7, but are significant in Band 9 data.  

\section{Residual maps} \label{appendix_c}

\begin{figure}[h!]
 \centering 
   \includegraphics[width=9.0cm]{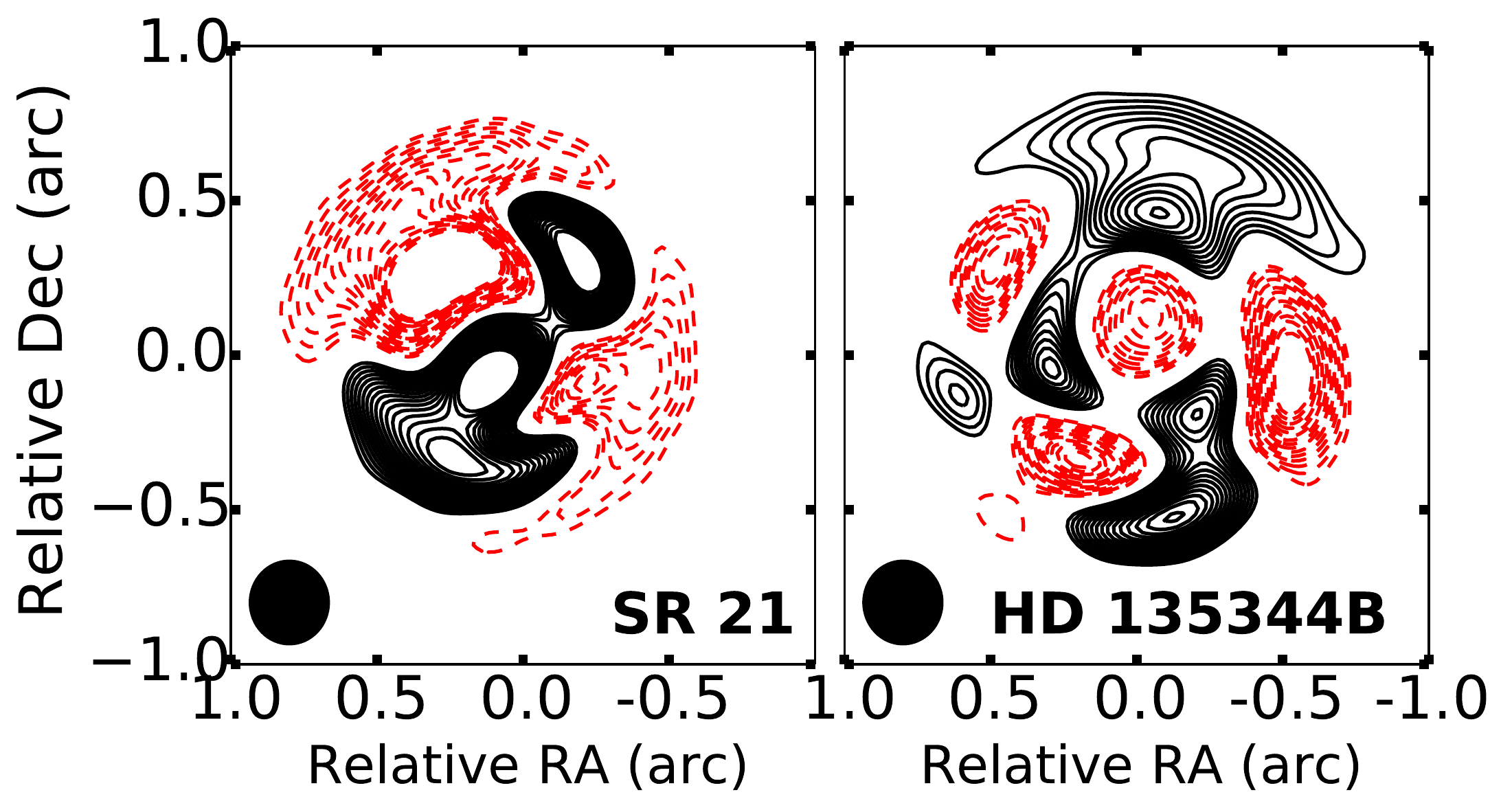}
   \caption{{\small Residual maps after subtracting the best-fit models from Table~\ref{best_parameters}, contours start at $\pm3\sigma$ and space by $3\sigma$ steps. }} 
   \label{residuals}
\end{figure}

Figure~\ref{residuals} illustrates the residual maps for both sources after subtracting the best-fit models (Table~\ref{best_parameters}) to the Band~7 data (336 GHz). For SR~21 the best-fit is described by a ring model, whereas for HD~135344B it is a vortex model. The residuals for HD~135344B show a spiral structure, as also suggested by \cite{perez2014} in Band~9 (689~GHz).

\section{Uncertainties in the azimuthal variations of the spectral index} \label{appendix_b}
\begin{figure}[h!]
 \centering 
   \includegraphics[width=8.0cm]{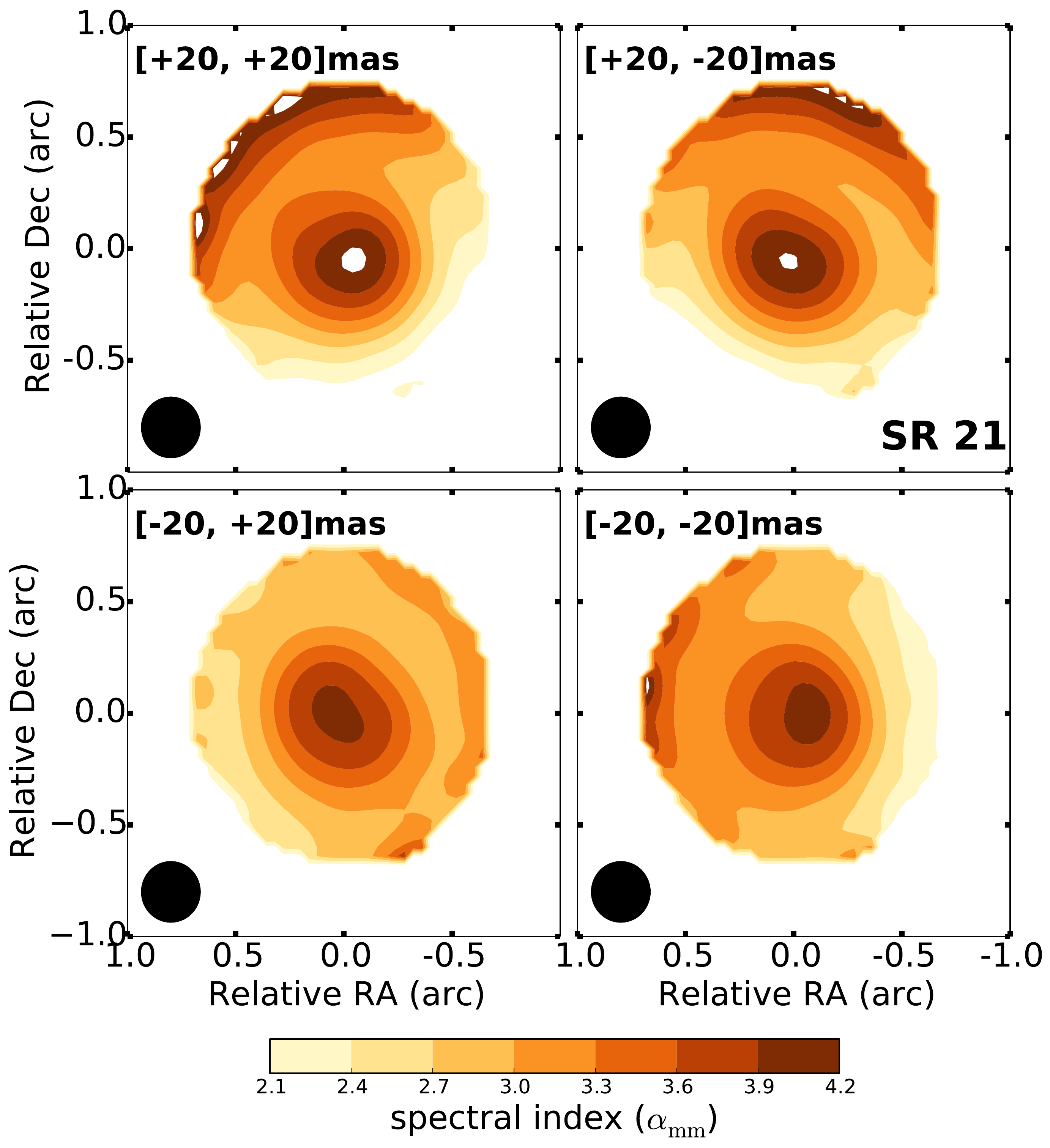}\\
   \includegraphics[width=8.0cm]{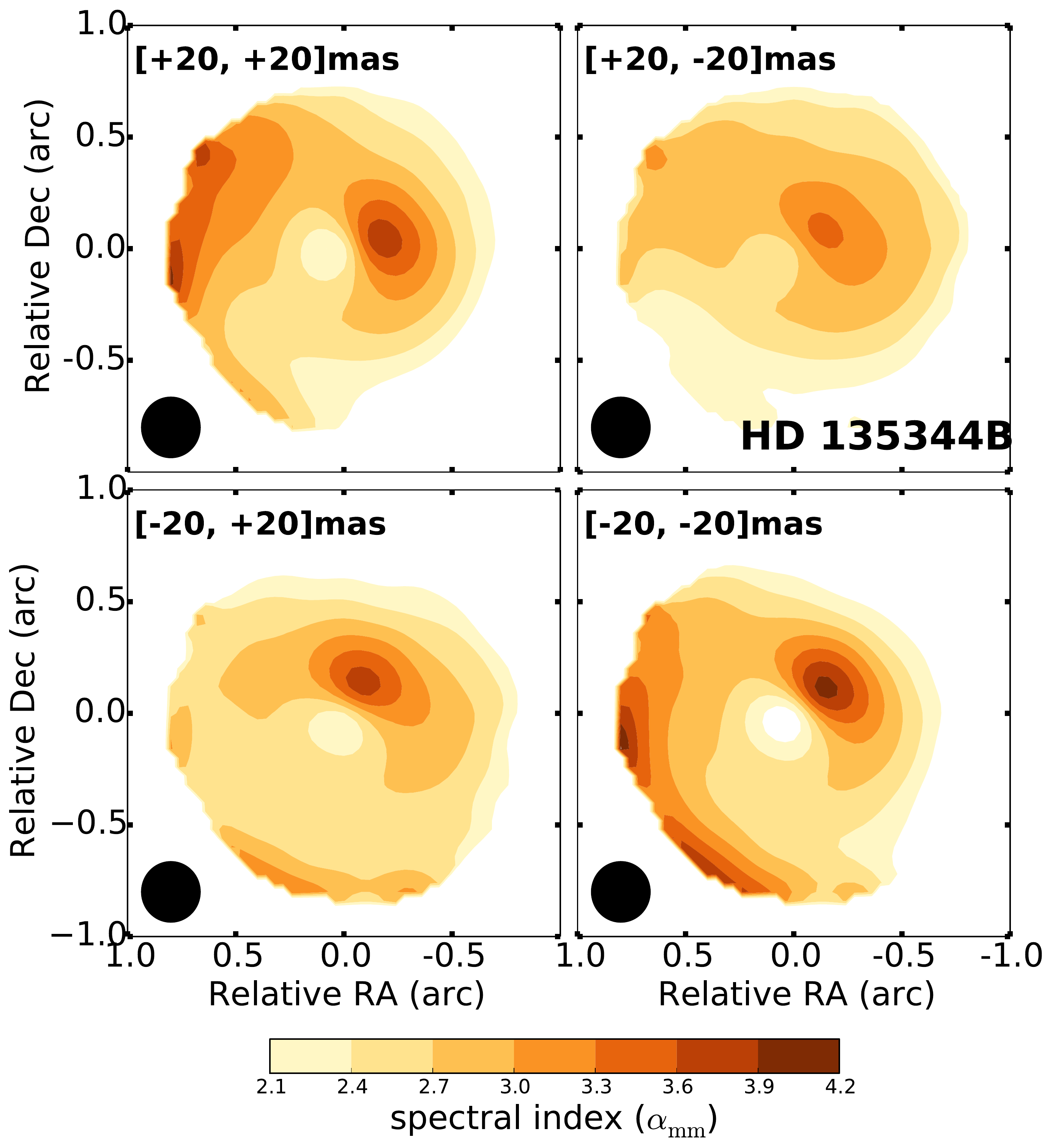}
   \caption{{\small Calculated spectral index ($\alpha_{\rm{mm}}$) assuming different alignments of the images for SR~21 (top panel) and HD~135344B (bottom panel).}}
   \label{shifts}
\end{figure}

The calculation of the spectral index ($\alpha_{\rm{mm}}$) from the observations depends considerably on how the two images are overlapped. To demonstrate that the apparent azimuthal variations remain uncertain, we shifted the alignment of the Band 7 and Band 9 images of SR21 and HD135344B by the positional uncertainty.  Figure~\ref{shifts} shows $\alpha_{\rm{mm}}$ by shifting 20~mas in both vertical and horizontal direction. The only reliable variation of $\alpha_{\rm{mm}}$  is in the radial direction for SR~21, which remains significant independent of the alignment. In  HD~135344B, there is a hint of high values of $\alpha_{\rm{mm}}$  opposite to the azimuthal asymmetry, which also remains independent of the alignment.

\end{document}